# Measurement of Nuchal Translucency Thickness for Detection of Chromosomal Abnormalities using First Trimester Ultrasound Fetal Images

S. Nirmala
Center for Advanced Research,
Muthayammal Engineering College,Rasipuram

.

V. Palanisamy
Info Institute of Engineering,
Kovilpalayam, Coimbatore – 641 107.

.

*Abstract*—The Nuchal Translucency thickness measurement is made to identify the Down Syndrome in screening first trimester fetus and presented in this paper. The mean shift analysis and canny operators are utilized for segmenting the nuchal translucency region and the exact thickness has been estimated using Blob analysis. It is observed from the results that the fetus in the 14th week of Gestation is expected to have a nuchal translucency thickness of 1.87±0.25mm.

## I. INTRODUCTION

In the recent past, the non invasive prenatal diagnoses of chromosomal disorders have been focused by researchers for detecting Down Syndrome (DS) fetuses. Down syndrome or Trisomy 21 is recognized as severe and common chromosomal abnormality occurring approximately once in every 800 to 1000 live births and the risk increase with the maternal age. It is found from the literature that the Down syndrome is a genetic condition most commonly caused by the extra number 21 chromosome. Affected babies are likely to suffer from severe mental disability and have a high chance of physical disabilities, affecting in particular the heart, gastrointestinal tract, eyes and ears.

Individuals with DS have a distinct craniofacial phenotype with a prominent forehead, small overall size of the craniofacial complex and underdevelopment of the fronto-naso maxillary region with missing or small nasal bones. It is also reported that Down syndrome causes Alzheimer's disease and a 15 to 20 times higher risk of leukimea. The literatures reveal that about 20% die by the age of five years due to cardiac problems [1],[2]. Based on the observations the nuchal translucency (NT) features in first trimester fetal images has been considered to be an important parameter for the detection of DS. Measurement of NT thickness has proved to be one of the most discriminating prenatal markers in screening for chromosomal abnormalities such as tirsomies 13,18 and 21[3].Accumulation of fluid in the nuchal region ,the nuchal translucency, can be shown in the ultrasound images for all fetuses during the first trimester[4].Even in presence of a normal karyotype, a bigger NT thickness is also associated with structural defects and genetic syndromes[5]. Furthermore ,an increased NT thickness(>2.5mm) between 10 and 14 weeks has also been associated with an increased risk of congenital heart and genetic syndrome[6],[7],[8].

Piotr Sieroszewski et al. [9] have evaluated the NT thickness sonographically and concluded that estimation of the same increased the predictive value of DS. Mehmet Tunc Canda and Namik Demi demonstrated with sonographic studies that increased NT thickness alone can detect 75% of fetuses with trisomy 21 [10] .The reported measurements so far have been made with the decision of skilled sonographers. As the NT thickness is of few millimeters, a small variation in the measurement made by the sonographer may lead to wrong diagnosis. The computer aided evaluation is expected to enhance the NT thickness measurement.

The computer aided measurement overcomes the problems inherent in manual measurements. Bernardino et al[11] proposed a semi automatic measurement system, which used the sobel operator to detect the border of NT. However well known edge detection techniques determine the location of an edge by local evaluation of a single image feature such as intensity or the intensity gradient. But no single image feature can provide reliable border measurement in fetal US images. Yu-Bu-lee presented a semi-automated detection procedure for measuring NT thickness based on Dynamic Programming (DP) [12] improved by a nonlinear anisotropic diffusion filtering which has involved preprocessing an image, defining a region of interest in order to reduce interference from the image boundary and to adapt to different fetal head positions and sizes of NT layer. The limitation of the proposed method is that it can be only applied when fetal position is a horizontal as possible.

Keeping the above facts, in the paper a semi-automatic computer aided algorithm to measure the Nuchal Translucency thickness for detecting the DS fetuses is described. The results obtained are promising and will support the physicians for better diagnosis in clinical pathology.

## II. METHODS

The Block diagram of the proposed image processing system is shown in Fig1. The various process involved in NT measurement is provided in this section. The fetus image is obtained from the ultrasound system and subjected to preprocessing for the extraction of features.

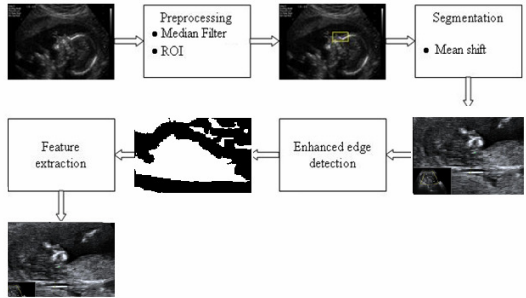

Figure. 1. Proposed Image Processing System.

### A. Image acquisition

The ultrasound fetus images are recorded by the ultrasound machine (model EMP 1100). The result is an exceptional precise beam providing enhancements in focus accuracy, spatial and contrast resolution. The probe used is a multifrequency probe of range 5-10 MHz. A perfect midsagittal view of the fetal profile is obtained. The probe is moved from side to side so that the inner edges of the two thin echogenic lines that border the NT layer is obtained. The magnification of the image should be such that the head and thorax region occupy a major portion of the image in the neutral position. The ultrasound images are obtained as the sequence of moving pictures. Still frame which is suitable for the proposed work is chosen.

### B. Image Characteristics

Nuchal translucency refers to the normal subcutaneous fluid-filled space between the back of the neck of a fetus and the overlying skin. Figure.2 shows a representative image of the NT and a schematic illustration of the echo zones (Z1-Z5) and the borders (B1-B2) of the NT layer. The NT thickness is defined as the maximum thickness of the translucent space (Z3) between the skin (Z4) and the soft tissues (Z2) overlying the cervical spine in the sagittal section of the fetus. Manual measurements of the NT layer are made by placing the crossbar of on-screen calipers on inner edges of the two thin echogenic lines that border the NT layer. Measurement of the maximum thickness should be made with the calipers perpendicular to the long axis of the fetus .An ultrasonographic sagittal view of a fetus and the calipers is shown in Figure 3a and Figure 3b illustrates correct and incorrect placement of the calipers on the anatomical structures of the fetus. [13]

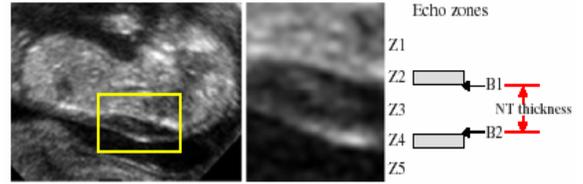

Figure. 2. Echo zones and NT borders

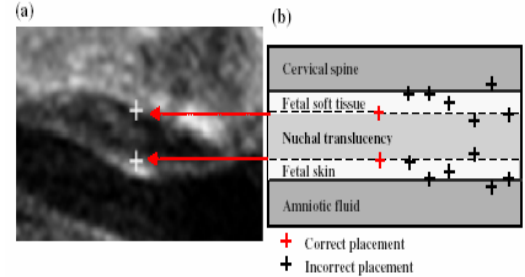

Figure. 3. Using calipers to measure NT thickness.
a) Ultrasonographic sagittal view of a fetus with calipers;
b) Placement of calipers on anatomical structures of thefetus

### C. Median Despeckling Filter

The Speckle noise commonly affects all coherent imaging systems and hence the images generated using Ultrasound technique appears to be much inferior to other imaging modalities [14]. Hence speckle filtering in diagnostic ultrasound provides the experts the enhanced diagnostic visibility. The speckle detection can be implemented with a median filter having a sliding window $S_w$ x $S_w$. If $S_w$ x $S_w$ is a window centered about $P_{(i, j)}$, then the filter coefficient of the median filter is given by

$$m_i^{n-1} = Med\left\{P_j^{n-1}, j \in \psi_i^{S_w}\right\} \tag{1}$$

Where P is the pixel to be analyzed and the neighbours are

$$\psi_i^{S_w} = \left\{P_i^{j+1}, P_{i+1}^{j+1}, P_{i+1}^{j}, P_i^{j-1}, P_{i-1}^{j}, P_{i-1}^{j+1}, P_{i-1}^{j-1}, P_{i+1}^{j-1}\right\} \tag{2}$$

The speckle detected sub image sequence 'Bi' is generated using the expression given below.

$$B_i^{n-1} = \left\{B_i^{n-1}, \Gamma < T_d\right\} \tag{3}$$

Where

$$\Gamma = m_i^{n-1} - \psi_i^{S_w} \tag{4}$$

And Td is the threshold value. If the value of Bj exceeds the threshold value, the co-efficient of second sub-image set or the decision making set is set to 1 or else it is set to 0. The second set is utilized for noise filtering process. Similar to the speckle detecting scheme, the noise filtering procedure also forms two sequences of sub-images. The first set of sub images can be denoted as {{Ni (0)}, {Ni (1)}, {Ni (2)}, {Ni (3)}….. {Ni (n)}}, is produced by low pass filtering at each Iteration levels. The second subset or the de-speckling subset, denoted by {{Si (0)}, {Si (1)}, {Si (2)}, {Si (3)}….. {Si(n)}}, is a binary set similar to that of the speckle detection scheme, where the good pixels are denoted by '1' and speckle corrupted pixels are marked as'0'. The procedure continues till all flag values in the binary sub-image sequence, {Bi (n)} is set to '1'. When all gray levels in {Bi (n)} set to '1', after the nth iteration, {Ii (n)} is the enhanced filter output.

*D. Region of Interest*

Fetal ultrasound images are expected to have wide variation because of the fetal movement during the scanning process. Therefore it is necessary to define a ROI which can compensate for changes in the fetal head position and NT region. As the main objective is to measure the thickness of NT, a suitable shape must be chosen for better diagnosis.

*E. Segmentation of NT Region*

The segmentation of US images is an essential component of computer assisted diagnosis system. The purpose of such systems is always to detect the boundaries of different organs from the diagnostic US images. Many segmentation methods have been proposed for medical imaging. The Mean shift algorithm has been found to be efficient in the segmentation of medical images due to its inherent advantages such as less noise, simple and provides better feature analysis [15, 16]. In the proposed work, the Mean shift analysis has been utilized for segmenting the nasal bone and the frontal bone. Any segment with pixels that share similar features will group together and may form clusters with a densely populated center in the feature space. The shift procedure detects clusters by determining modes in a parametric density function iteratively. The feature vector consists of the pixel position and gray value. Let {Xi} i=1,2,.. n be an arbitrary set of n points in the d dimensional Euclidian space $R_d$. The multivariate kernel density function for a set of points of {Xi} i=1…n is given by:

$$f'(x) = \frac{1}{nh^d} \sum_{i=1}^{n} k\{\frac{1}{h}(x - x_i)\} \qquad (5)$$

The dimension and h is the radius of the hyper sphere. K is the Epanechikov kernel defined as:

$$K_E(x) = \begin{cases} \frac{1}{2} C_d^{-1}(d+2)(1 - \|X\|^2), if \|X\| < 1 \\ 0, otherwise \end{cases} \qquad (6)$$

This yields the lowest Mean Squared Error (MISE) in the estimates computed. The gradient estimate $\Delta f(x)$ is obtained by substituting a differentiable kernel into $f'(x)$.

$$\Delta f'_E(X) = \frac{1}{n(h^d c_d)} \frac{d+2}{h^2} \sum_{X_i \in S_h(x)} (X_i - X)$$

$$= \frac{n_x}{n(h^d c_d)} \frac{d+2}{h^2} [\frac{1}{n_x} \sum_{X_i \in S_h(x)} (X_i - X)] - -(7)$$

$S_h(x)$ is a hyper sphere with volume $h^d c_d$ centered at X. $n_x$ is the total number of data points contained inside the hyper sphere.

The following equation is the Mean Shift Vector.

$$M_{h,U(X)} = \frac{1}{n_x} \sum_{X_x \in S_h(x)} (X_i - X) - -(8)$$

Convergence: The image format is converted into spatial feature space. An arbitrary number of hyper spheres are defined with centers {X₁, X₂, X₃, X_C} within the sample set. The locations of the center of the hyper spheres are updated iteratively according to:

$$X_f(n+1) = X_{fn} + M_{fh,U(x)} \qquad (9)$$

where the f subscript is the sphere number.

**Algorithm**

Let {xj} j=1…n be the original data points, {zj} j=1….n the points of convergence, and {Lj} j=1…n a set of labels. i.e. scalars.

1. For each j=1…n run the mean shift procedure for xj and store the convergence point in zj.
2. Identify clusters {Cp} p=1…m of convergence points by linking together all zj which are closer than 0.5 from each other in the joint domain.
3. For each j=1…n assign class labels Lj= {p | zj ε Cp} to clusters.
4. If desired, eliminate regions smaller than P pixels.

*F. Edge Detection*

The segmented image is then subjected to canny edge detection for enhanced visibility of edges for further processing of the image.

III. Results and Discussion

Experiments have been carried out on a set of sonographic images obtained from the Ultrasound machine. A total of 50 images under various weeks of gestation were considered for analysis. The images were preprocessed and the region of interest has been cropped out for further analysis .NT region has been segmented from the cropped image by applying the mean shift cluster analysis. The data at the edges has been enhanced using canny operator to improve the visibility of the data. The NT thickness has been estimated using Blob analysis. The implementation steps were as follows.

Step1:  Image acquisition using Ultrasound system.
Step2:  Removal of Speckle noise using Median filter.
Step3:  Identification of NT (ROI).
Step4:  Applying Mean shift segmentation Algorithm for segmenting the ROI.
Step5:  Edge Detection of the segmented NT by Canny edge detection.
Step6:  Binary masking of the entire image
Step7:  Blob analysis for the measurement of thickness of Nuchal Translucency.

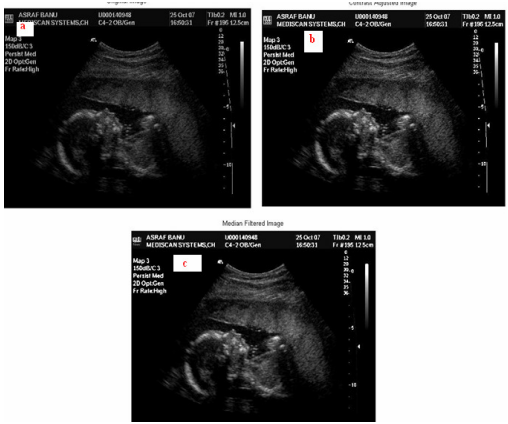

Figure. 4. Sample and Filtering Image.

The sample image and the despeckled image after preprocessing is shown in Figure 4.

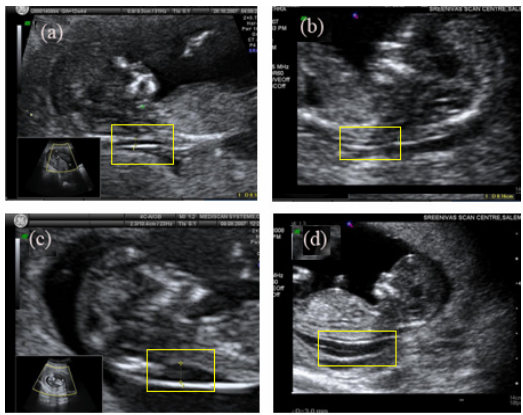

Figure. 5. Selection of ROI

The region of interest has been obtained for further segmentation. Figure.5 shows different views of ROI of fetal images. The selection of ROI will lead to precise segmentation.

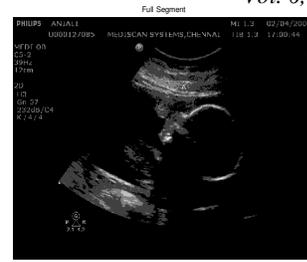

Figure. 6 (a). Segmented image of normal fetus

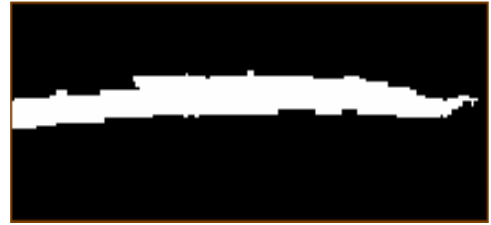

Figure. 6 (b). Edge detected output of normal fetus

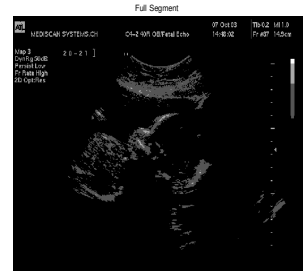

Figure. 6 (c). Segmented output of abnormal fetus

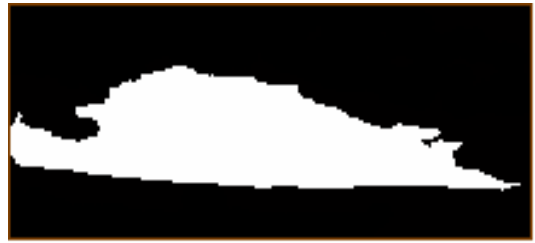

Figure. 6 (d). Edge detected output of abnormal fetus

The segmented image and edge detected image are shown in Figure 6.The Mean shift algorithm is utilized for obtaining the segmentation of NT region. For normal and abnormal fetus images, the visibility of the edges are improved using canny operator. The boundary of the segmented NT region is shown in Figure 6(b) and 6(d) for normal and anueploid fetus respectively.It is obvious from the results that the abnormal fetus has enlarged NT thickness when compared with the normal fetus.

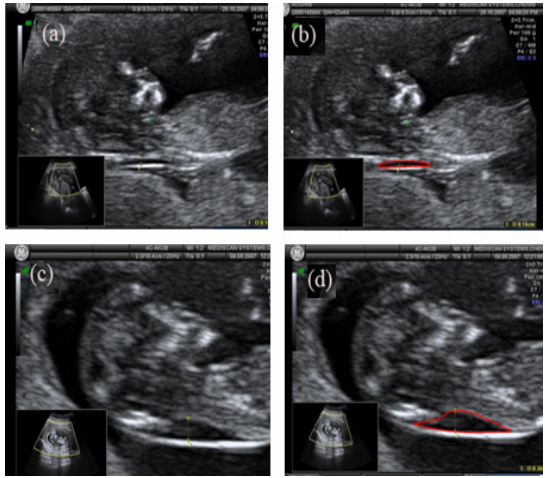

Figure. 7. Image of various fetus for different weeks of gestation

Figure 7(a) and 7(c) depict the normal and abnormal fetus image of different weeks of gestation. In Figure 7(b) and 7(d) the extracted NT region is superimposed on the original image and marked as red. It is made as a verification methodology for ensuring the segmentation process.

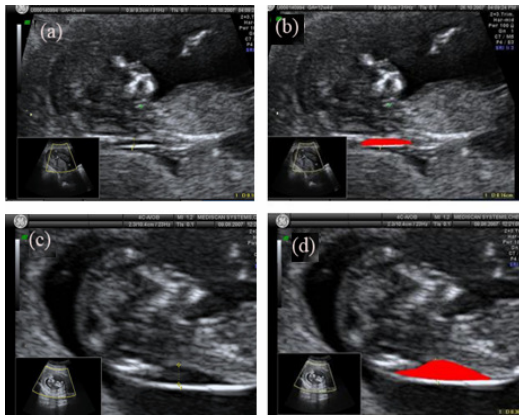

Figure. 8. Images of fetus for different weeks of gestation with NT region Superimposed

TABLE 1. NT THICKNESS IN FIRST TRIMESTER MEASURED FOR NORMAL FETUS WITH GESTATIONAL PERIOD VARYING FROM 11 TO 14 WEEKS

| S.no | Gestation Weeks | Number of subjects | Average (mm) $x$ | Standard Deviation $\sigma$ | Variance $\sigma^2$ |
|------|------|------|------|------|------|
| 1 | 11/11.6 | 11 | 1.35±0.41 | 0.114 | 0.0145 |
| 2 | 12/12.6 | 10 | 1.45±0.43 | 0.223 | 0.0595 |
| 3 | 13/13.6 | 12 | 1.11±0.62 | 0.215 | 0.0337 |
| 4 | 14/14.6 | 10 | 1.87±0.25 | 0.117 | 0.0183 |

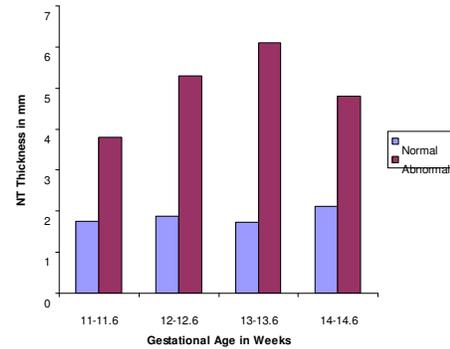

Figure. 9. Plot of mean values of NT for various weeks of gestation

The estimated NT thickness for the total population including the normal and abnormal fetus has been graphically shown in Figure 9. It can be observed from the graphical results that the normal fetuses have an average Nuchal translucency thickness which is considerably less compared to the abnormal fetuses.

It is also observed that the computer aided measurement enhances the estimation. The deviation of the Nuchal Translucency thickness from the average thickness measured for the set of samples indicates that the measurement is more reliable. For example the fetus in 14 weeks of gestation is expected to have the NT thickness of 1.87 ± 0.25 mm.

IV CONCLUSION

The measurements of NT for normal and abnormal fetus have been carried out. It is concluded that the computed aided measurement will provide valuable information to the physicians to take accurate decision. The results reveal that the normal fetus with gestation week of 14 must not have greater than 2.12 mm of NT thickness

ACKNOWLEDGMENT

The authors would like to thank Dr.M.Madheswaran, Principal, Muthayammal Engineering College for his valuable suggestions. They would like to express their gratitude to Dr. S.Suresh, Mediscan Systems, Chennai for providing the necessary images for this study.

AUTHORS PROFILE

**S.Nirmala** received her B.E in Electrical and Electronics Engineering from Government College of Engineering, Salem, Madras University and Master Degree in Applied Electronics from Anna University, Chennai. She is doing her Ph.D in Biomedical Imaging at Anna University, Chennai. She is at present working as an Assistant Professor in the Department of Electronics and Communication Engineering, Muthayammal Engineering College, Rasipuram, Namakkal District, Tamilnadu. Her field of interest is Biomedical Engineering, Image Processing, Digital Signal Processing, VLSI Design and Embedded systems.She is a member of IEEE and vice chairman of IEEE India CAS chapter.

**Dr.V.Palanisamy** received his B.E degree in Electronics and Communication Engineering from PSG College of Technology, Coimbatore and Master Degree in Communication systems from University of Madras. He also received his Ph.D in Antennas Theory from Indian Institute of Technology, Karagpur. Since 1974 he has been working in various capacities in the Department of Technical Education in Tamilnadu. He is at present working as a Principal of Info Institute of Engineering, Coimbatore. His field of interest is Electronics, Antennas, Image Processing, Communication Systems and VLSI Technologies.